\documentclass[aps,preprint]{revtex4}

\newcommand{\BE}{\begin{equation}}
\newcommand{\EE}{\end{equation}}
\newcommand{\BA}{\begin{eqnarray}}
\newcommand{\EA}{\end{eqnarray}}

\begin{document}

\title{Theoretical analysis of acoustic stop bands in two-dimensional periodic scattering arrays}
\author{You-Yu Chen and Zhen Ye}
\affiliation{Wave Phenomena Laboratory,~Department of
Physics,~National Central University, Chungli, Taiwan 32054\ \\ \
\\ \ }


\begin{abstract}

This paper presents a theoretical analysis of the recently
reported observation of acoustic stop bands in two-dimensional
scattering arrays (Robertson and Rudy, J. Acoust. Soc. Am. {\bf
104}, 694, 1998). A self-consistent wave scattering theory,
incorporating all orders of multiple scattering, is used to obtain
the wave transmission. The band structures for the regular arrays
of cylinders are computed using the plane wave expansion method.
The theoretical results compare favorably with the experimental
data. (PACS numbers: 43.20.Gp, 43.20.Ye, 43.20.Fn; April 14, 2001)

\end{abstract}

\maketitle

\section{Introduction}

When propagating through media containing many scatterers, waves
will be scattered by each scatterer. The scattered waves will be
scattered again by other scatterers. This process is repeated to
establish an infinite iterative pattern of rescattering between
scatterers, forming a multiple scattering process \cite{Ishimaru}.
Multiple scattering of waves is responsible for a wide range of
fascinating phenomena, including such as twinkling light in the
evening sky, modulation of ocean ambient sound\cite{Ye1}. On
smaller scales, phenomena such as white paint, random
laser\cite{Laser}, electron transport in impured solids\cite{Im}
are also results of multiple scattering. When waves propagate
through media with periodic structures, the multiple scattering
leads to the phenomenon of band structures. That is, waves can
propagate in certain frequency ranges and follow certain
dispersion relations, while in other frequency regimes wave
propagation may be stopped. The former ranges are called allowed
bands and the latter the forbidden bands.

The wave dispersion bands were first studied for electronic waves
in solids, providing the basis for understanding the properties of
conductors, semi-conductors, and insulators\cite{Kittel}. In late
1980s, it became known that such a wave band phenomenon is also
possible for classical waves. The studies on manipulation of
classical waves were started with electro-magnetic waves in media
with periodically modulated refractive-indices\cite{Optic}. Since
then, optical wave bands have been extensively studied, yielding a
rich body of literature\cite{web}. The theoretical calculations
have proven to match well with the experimental
observations\cite{Exp1}. The modulation of optical waves by
periodic media has led to a number of practical applications
including the design of photonic crystals\cite{pc}, the effective
optical fibers\cite{fiber} and waveguide devices\cite{waveguide}.
Recently, it has also been found that a living organism may also
display a remarkable photonic engineering\cite{seamouse}.

In contrast, research on acoustic wave band structures has just
started (For example, refer to \cite{1998}). Although theoretical
computations of band structures have been well documented for
periodic acoustic structures\cite{Kush}, the experimental work was
only recent, and to date only a limited number of measurements
have been reported. One of the first observations was made on
acoustic attenuation by a minimalist sculpture\cite{Sculpture} and
further studied in the laboratory\cite{Sanchez}. The authors
obtained a sound attenuation spectrum, which was later verified by
the band structure computation\cite{Kush2,chen}. Recently,
acoustic band structures have been further measured for acoustic
transmission through two-dimensional (2D) periodic arrays of metal
cylinders placed in the air\cite{1998}. The authors reported
experimental observation of acoustic stop bands and wave
transmission for both square and triangular arrays. The impulse
response technique was used to determine the transmission over a
broad frequency bandwith, whereas the acoustic dispersion relation
was extracted from the phase information.

The main purpose of this paper is to provide a theoretical
investigation of sound transmission by 2D arrays of rigid
cylinders in air in line with the experiment described by
Robertson and Rudy\cite{1998}, providing a direct comparison of
the acoustic transmission between theory and experiment. For the
purpose, we employ a self-consistent multiple scattering
theory\cite{Twersky} to compute the acoustic transmission through
arrays of scattering cylinders. Meanwhile, the acoustic band
structures are computed using the plane-wave method well
prescribed by Kushwaha\cite{Kush}. We will show that the
theoretical results agree very well with the observation.

\section{Formulation of the problem}

\subsection{Acoustic scattering by arrays of parallel cylinders}

Consider $N$ straight identical cylinders located  at $\vec{r}_i$
with $i=1, 2, \cdots, N$ to form either a regular lattice (or a
random array) perpendicular to the $x-y$ plane; the regular
arrangement can be adjusted to comply with the
experiment\cite{1998}. There are two types of the regular
arrangements of the cylinders: the square lattice and the
triangular lattice. The cylinders are along the $z$-axis. An
acoustic source transmitting monochromatic waves is placed at
$\vec{r}_s$, some distance from the array. The scattered wave from
each cylinder is a response to the total incident wave composed of
the direct wave from the source and the multiply scattered waves
from other cylinders. The final wave reaches a receiver located at
$\vec{r}_r$ is the sum of the direct wave from the source and the
scattered waves from all the cylinders. The cylinders used in the
experiment is metal cylinders. Numerical computation verifies that
for acoustic scattering, the effect due to the shear modulus is
negligible for such a cylinder in air. This has also been
confirmed by experiments\cite{Sanchez}. When the shear waves are
ignored, the exact solution for the scattering process can be
conveniently formulated, following Twersky\cite{Twersky}.

To exactly reproduce the experimental data, it would have to know
the information about the apparatus, the acoustic pulses
generated, the lab environment, and the arrangement of the
sounding and receiving devices. As the information is not readily
available and is also unnecessary for the present theoretical
investigation, we make certain reasonable simplifications.

For simplicity yet without compromising generality, we approximate
the acoustic source as a line source located at origin, i.~e.
$\vec{r}_s =0$; the numerical computation indicates that the
difference between a line source and a plane wave is not
essential. Without the cylinders, the wave is governed by \BE
(\nabla^2 +k^2)G(\vec{r}) = -4\pi\delta^{(2)}(\vec{r}), \EE where
$H_0^{(1)}$ is the zero-th order Hankel function of the first
kind. In the cylindrical coordinates, the solution is \BE
G(\vec{r}) = \mbox{i}\pi H_0^{(1)}(kr). \EE In this section,
`$\mbox{i}$' stands for $\sqrt{-1}$.

With $N$ cylinders located at $\vec{r}_i$ ($i=1,2,\cdots, N$), the
scattered wave from the $j$-th cylinder can be written as \BE
\label{eqps1} p_s(\vec{r}, \vec{r}_j) = \sum_{n=-\infty}^{\infty}
\mbox{i}\pi A_n^j H_n^{(1)}(k|\vec{r} -
\vec{r}_j|)e^{\mbox{i}n\phi_{\vec{r}- \vec{r}_j}}, \EE where
$H_n^{(1)}$ is the $n$-th order Hankel function of the first kind.
$A_n^i$ is the coefficient to be determined, and $\phi_{\vec{r}-
\vec{r}_j}$ is the azimuthal angle of the vector $\vec{r}-
\vec{r}_i$ relative to the positive $x$-axis.

The total wave incident around the $i$-th scatterer
$p_{in}^i(\vec{r})$ is a superposition of the direct contribution
from the source $p_0(\vec{r}) = G(\vec{r})$ and the scattered
waves from all other scatterers: \BE \label{eqpin1}
p_{in}^i(\vec{r}) = p_0(\vec{r}) + \sum_{j=1,j\neq i}^N
p_s(\vec{r}, \vec{r}_j). \EE In order to seperate the governing
equations into modes, we can express the total incident wave in
term of the modes about $\vec{r_i}$: \BE \label{eqpin2}
p_{in}^i(\vec{r}) = \sum_{n = -\infty}^\infty B_n^i J_n(k|\vec{r}
- \vec{r_i}|)e^{\mbox{i}n\phi_{\vec{r} - \vec{r_i}}}. \EE The
expansion is in terms of Bessel functions of the first kind $J_n$
to ensure that $p_{in}^i(\vec{r})$ does not diverge as $\vec{r}
\rightarrow \vec{r_i}$.  The coefficients $B_n^i$ are related to
the $A_n^j$ in equation (\ref{eqps1}) through equation
(\ref{eqpin1}).  A particular $B_n^i$ represents the strength of
the $n$-th mode of the total incident wave on the $i$-th scatterer
with respect to the $i$-th scatterer's coordinate system (i.e.
around $\vec{r_i}$).  In order to isolate this mode on the right
hand side of equation (\ref{eqpin1}), and thus determine a
particular $B_n^i$ in terms of the set of $A_n^j$, we need to
express $p_s(\vec{r}, \vec{r_j})$, for each $j \neq i$, in terms
of the modes with respect to the $i$-th scatterer. In other words,
we want $p_s(\vec{r}, \vec{r_j})$ in the form \BE \label{eqps2}
p_s(\vec{r}, \vec{r_j}) = \sum_{n=-\infty}^\infty C_n^{j, i}
J_n(k|\vec{r} - \vec{r_i}|)e^{\mbox{i}\phi_{\vec{r} - \vec{r_i}}}.
\EE This can be acheived (i.e. $C_n^{j, i}$ expressed in terms of
$A_n^i$) through the following addition theorem\cite{addition}:
\BE \label{eqaddition} H_n^{(1)}(k|\vec{r} -
\vec{r_j}|)e^{\mbox{i}n\phi_{\vec{r} - \vec{r_j}}} =
e^{\mbox{i}n\phi_{\vec{r_i} - \vec{r_j}}}
\sum_{l=-\infty}^{\infty} H_{n-l}^{(1)}(k|\vec{r_i} - \vec{r_j}|)
e^{-\mbox{i}l\phi_{\vec{r_i} - \vec{r_j}}} J_l(k|\vec{r} -
\vec{r_i}|) e^{\mbox{i}l\phi_{\vec{r} - \vec{r_i}}}. \EE

Taking equation (\ref{eqaddition}) into equation (\ref{eqps1}), we
have \BE p_s(\vec{r}, \vec{r_j}) = \sum_{n=-\infty}^\infty
\mbox{i}\pi A_n^j e^{\mbox{i}n\phi_{\vec{r_i} - \vec{r_j}}}
\sum_{l=-\infty}^{\infty} H_{n-l}^{(1)}(k|\vec{r_i} - \vec{r_j}|)
e^{-\mbox{i}l\phi_{\vec{r_i} - \vec{r_j}}} J_l(k|\vec{r} -
\vec{r_i}|) e^{\mbox{i}l\phi_{\vec{r} - \vec{r_i}}}. \EE Or by
switching the order of summation, we have \BE p_s(\vec{r},
\vec{r_j}) = \sum_{l=-\infty}^{\infty}
\left[\sum_{n=-\infty}^\infty \mbox{i}\pi A_n^j
H_{n-l}^{(1)}(k|\vec{r_i} - \vec{r_j}|)
e^{\mbox{i}(n-l)\phi_{\vec{r_i} - \vec{r_j}}} \right]
J_l(k|\vec{r} - \vec{r_i}|) e^{\mbox{i}l\phi_{\vec{r} -
\vec{r_i}}}. \EE Comparing with equation (\ref{eqps2}), we see
that \BE C_n^{j,i} = \sum_{l=-\infty}^\infty \mbox{i}\pi A_l^j
H_{l-n}^{(1)}(k|\vec{r_i} - \vec{r_j}|)
e^{\mbox{i}(l-n)\phi_{\vec{r_i} - \vec{r_j}}} \EE

Now we can relate $B_n^i$ to $C_n^{j, i}$ (and thus to $A_l^j$)
through equation (\ref{eqpin1}).  First note that through the
addition theorem the source wave can be written, \BE
\label{eqp0exp}
\begin{array}{lll}
p_0(\vec{r}) & = & \mbox{i}\pi H_0^{(1)}(kr) \\ & = & \mbox{i}\pi
\sum_{l=-\infty}^{\infty} H_{-l}^{(1)}(k|\vec{r_i}|)
e^{-\mbox{i}l\phi_{\vec{r_i}}} J_l(k|\vec{r} - \vec{r_i}|)
e^{\mbox{i}l\phi_{\vec{r} - \vec{r_i}}} \\ & = &
\sum_{l=-\infty}^{\infty} S_l^i J_l(k|\vec{r} - \vec{r_i}|)
e^{\mbox{i}l\phi_{\vec{r} - \vec{r_i}}},
\end{array}
\EE where \BE S_l^i = \mbox{i}\pi H_{-l}^{(1)}(k|\vec{r_i}|)
e^{-\mbox{i}l\phi_{\vec{r_i}}}. \EE Matching coefficients in
equation (\ref{eqpin1}) and using equations (\ref{eqpin2}),
(\ref{eqps2}) and (\ref{eqp0exp}), we have \BE B_n^i = S_n^i +
\sum_{j=1,j\neq i}^N C_n^{j, i}, \EE or, expanding $C_n^{j, i}$,
\BE \label{eqmatrix1} B_n^i = S_n^i + \sum_{j=1,j\neq i}^N
\sum_{l=-\infty}^\infty \mbox{i}\pi A_l^j
H_{l-n}^{(1)}(k|\vec{r_i} - \vec{r_j}|)
e^{\mbox{i}(l-n)\phi_{\vec{r_i} - \vec{r_j}}}. \EE At this stage,
both the $S_n^i$ are known, but both $B_n^i$ and $A_l^j$ are
unknown.  Boundary conditions will give another equation relating
them.

The boundary conditions are that the pressure and the normal
velocity be continuous across the interface between a scatterer
and the surrounding medium.  The total wave outside the $i$-th
scatterer is $p_{ext} = p_{in}^i(\vec{r}) + p_s(\vec{r},
\vec{r_i})$.  The wave inside the $i$-th scatterer can be
expressed as \BE \label{eqpint1} p_{int}^i(\vec{r}) = \sum_{n =
-\infty}^\infty D_n^i J_n(k_1^i|\vec{r} - \vec{r_i}|)
e^{\mbox{i}n\phi_{\vec{r} - \vec{r_i}}}. \EE The boundary
conditions are then \BE p_{ext}|_{\partial\Omega^i} =
p_{int}|_{\partial\Omega^i} \EE and \BE
\frac{1}{\rho}\left.\frac{\partial p_{ext}}{\partial n}\right
|_{\partial\Omega^i} = \frac{1}{\rho_1^i}\left.\frac{\partial
 p_{int}}{\partial n}\right |_{\partial\Omega^i},
\EE where $\partial\Omega^i$ is the boundary of the $i$-th
scatterer, $k$ and $\rho$ are the wavenumber and density of the
surrounding medium, and $k_1^i$ and $\rho_1^i$ are the wavenumber
and density of the $i$-th scatterer respectively. Using equations
(\ref{eqps1}), (\ref{eqpin2}) and (\ref{eqpint1}), multiplying
both sides of the boundary condition equations by
$e^{\mbox{i}n\phi_{\vec{r} - \vec{r_i}}}$, and integrating over
the boundary $\partial\Omega^i$, we have for the case of circular
cylindrical scatterers, \BA B_n^i J_n(k a^i) + \mbox{i}\pi A_n^i
H_n^{(1)}(k a^i) & = & D_n^i J_n(k a^i / h^i) \\ B_n^i J_n'(k a^i)
+ \mbox{i}\pi A_n^i H_n^{(1)\prime}(k a^i) & = & \frac{1}{g^i h^i}
D_n^i J_n'(k a^i / h^i). \EA Here $a^i$ is the radius of the
$i$-th cylinder, $g^i = \rho_1^i/\rho$ is the density ratio, and
$h^i = k/k_1^i = c_1^i/c$ is the sound speed ratio for the $i$-th
cylinder.  Elimination of $D_n^i$ gives \BE B_n^i =
\mbox{i}\pi\Gamma_n^i A_n^i, \EE where \BE \Gamma_n^i = \left[
\frac{H_n^{(1)}(k a^i) J_n'(k a^i/h^i) - g^i h^i H_n^{(1)\prime}(k
a^i) J_n(k a^i/h^i)}
     {g^i h^i J_n'(k a^i) J_n(k a^i/h^i) - J_n(k a^i) J_n'(k a^i/h^i)}
\right] \EE

If we define \BE T_n^i = S_n^i/\mbox{i}\pi =
H_{-n}^{(1)}(k|\vec{r_i}|) e^{-\mbox{i}n\phi_{\vec{r_i}}} \EE and
\BE G_{l,n}^{i,j} = H_{l-n}^{(1)}(k|\vec{r_i} - \vec{r_j}|)
e^{\mbox{i}(l-n)\phi_{\vec{r_i} - \vec{r_j}}}, i\neq j \EE then
equation (\ref{eqmatrix1}) becomes \BE \label{eqfinalmatrix}
\Gamma_n^i A_n^i - \sum_{j=1,j\neq i}^N \sum_{l=-\infty}^\infty
G_{l,n}^{i,j} A_l^j = T_n^i. \EE If the value of $n$ is limited to
some finite range, then this is a matrix equation for the
coefficients $A_n^i$.  Once solved, the total wave at any point
outside all cylinders is \BE p(\vec{r}) = \mbox{i}\pi
H_0^{(1)}(k|\vec{r}|) + \sum_{i=1}^N \sum_{n=-\infty}^\infty
\mbox{i}\pi A_n^i H_n^{(1)}(k|\vec{r} - \vec{r_i}|)
e^{\mbox{i}n\phi_{\vec{r} - \vec{r_i}}}.\label{eq:final} \EE We
must stress that total wave expressed by eq.~(\ref{eq:final})
incorporate all orders of multiple scattering. We note, however,
that an inclusion of the lowest order in multiple scattering may
be sufficient for certain situations (J. S\'anchez-Dehesa, private
communication). We also emphasize that the above derivation is
valid for any configuration of the cylinders. In other words,
eq.~(\ref{eq:final}) works for situations that the cylinders can
be placed either randomly or orderly.

\subsection{Band structures of regular arrays of cylinders}

For a regular array of the cylinders, band structures for the wave
propagation appear. The band structures can be readily computed by
the plane-wave method\cite{Kush}. Though the method has been well
documented by Kushwaha\cite{Kush}, for the sake of convenience we
outline the approach as follows.

The wave equation is \BE
\nabla\cdot\left[\frac{1}{\rho(\vec{r})}\nabla p(\vec{r})\right] +
\frac{\omega^2}{\rho(\vec{r})c^2(\vec{r})} p(\vec{r})=0,
\label{eq:wave}\EE where $\rho(\vec{r})$ and $c(\vec{r})$ are the
mass density and sound speed respectively; both are modulated by
the periodic structures, i.~e. inside the cylinders the values are
that of the cylinders, while outside the cylinders they take the
values of the medium. According to Bloch's theorem\cite{Kittel},
the solution of the pressure field has the Bloch form \BE
p(\vec{r}) =
e^{i\vec{K}\cdot\vec{r}}\sum_{\vec{G}}\phi_{\vec{K}}(\vec{G})e^{i\vec{G}\cdot\vec{r}},
\label{eq:bloch}\EE where $\vec{K}$ is termed as the Bloch
wavevector, $\vec{G}$ is the reciprocal lattice
vector\cite{Kittel}. The summation is made for all possible
reciprocal vectors.

For the periodic structures, both $\rho^{-1}$ and $(\rho
c^2)^{-1}$ in eq.~(\ref{eq:wave}) can be expanded by discrete
plane waves as follows \BE \frac{1}{\rho(\vec{r})} =
\sum_{\vec{G}}\sigma(\vec{G}) e^{i\vec{G}\cdot\vec{r}}, \
\mbox{and} \ \frac{1}{\rho(\vec{r})c^2(\vec{r})} = \sum_{\vec{G}}
\eta(\vec{G})e^{i\vec{G}\cdot\vec{r}}. \label{eq:exp}\EE As
$\rho(\vec{r})$ and $c(\vec{r})$ are known parameters, both
$\sigma(\vec{G})$ and $\eta(\vec{G})$ can be determined from a
inverse Fourier transform.

Substituting eqs.~(\ref{eq:bloch}) and (\ref{eq:exp}) into
eq.~(\ref{eq:wave}), we obtain \BE
\sum_{\vec{G}'}\left[\sigma(\vec{G}-\vec{G}')(\vec{K} +
\vec{G})\cdot(\vec{K} + \vec{G}') -
\eta(\vec{G}-\vec{G}')\omega^2\right] \phi_{\vec{K}}(\vec{G}')=0,
\EE which has the matrix form $$
\sum_{\vec{G}'}\Gamma_{\vec{G},\vec{G}'} \phi_{\vec{K}}(\vec{G}')
= 0. $$ The dispersion relation connecting the frequency $\omega$
and the wave vector $\vec{K}$ is determined by the secular
equation \BE \mbox{det}[\Gamma_{\vec{G},\vec{G}'}] =
\mbox{det}\left[\sigma(\vec{G}-\vec{G}')(\vec{K} +
\vec{G})\cdot(\vec{K} + \vec{G}') -
\eta(\vec{G}-\vec{G}')\omega^2\right]_{\vec{G},\vec{G}'} = 0,
\label{eq:dis}\EE where `det' denotes the determinant.
Eq.~(\ref{eq:dis}) leads to the dispersion relation between the
frequency $\omega(\vec{K})$ and the wave vector $\vec{K}$.

\section{Numerical results}

Numerical computation has been performed to obtain the transmitted
acoustic wave and the acoustic band structures. In particular, the
numerical computation has been carried out for the experimental
situations\cite{1998}.

First we consider the transmitted waves described by
eq.~(\ref{eq:final}). In the simulation, all the cylinders are
assumed to be the same, in accordance with the experiment.
Moreover, the radii of the cylinders and the lattice constants are
also taken from the experiment. Several values for the acoustic
contrasts between the cylinder and the air and different cylinders
including the conduit cylinders originally used in the experiment
were used in the initial stage of computation. We found that the
results are in fact insensitive to the detailed material
composition of the cylinders as long as the contrasts exceed a
certain value. This agrees with the previous experimental
observation\cite{Sanchez} and the theoretical results\cite{chen}.
In the present computation, we also allow the total number of the
cylinders to vary from 36 to 500. The cylinders are placed to form
a square lattice or a triangular lattice.

Assume that the lattice spacing is $a$, the diameter of the
cylinders is $d$. For the square lattice, the filling factor, that
is the fraction of the sample area occupied by the scattering
cylinders, is calculated as \BE f = \frac{\pi d^2}{4a^2}.\EE For
the triangular lattice, the filling factor is given as \BE f =
\frac{\pi d^2}{2\sqrt{3} a^2}.\label{eq:2}\EE Here we note an
editorial error in eq.~(2) of \cite{1998}. In the experiment
carried out in \cite{1998}, the following parameters are used. (1)
For the square lattice, $a=3.7$ cm and $d=2.34$ cm. This gives a
filling factor of 0.31. (2) For the triangular lattice, for the
same cylinders, the filling factor is fixed as $0.366$, leading to
a spacing of 3.683 cm.

The theoretical results show that the wave transmission is
sensitive to the filling factor, as well as the number of the
cylinders. In order to limit the possible finite size effects so
that they do not obscure the observation of the band gap effects,
we found that we need to have more rods than used in the
experiment. For frequencies at which wave propagation is possible,
there is sensitive interference between the propagating wave and
the reflected waves at the boundaries, yielding the familiar
pattern of nulls and peaks. If there is a band gap, the
transmission will not be possible within the gap. Then the
received signal is small, and the transmission will be relatively
insensitive to the boundary effects. In other words, the
inhibition of a band gap will not be altered by varying the sample
size. Our numerical results confirm this.

\begin{figure}[hbt]
\begin{center}
\epsfxsize=5in \epsffile{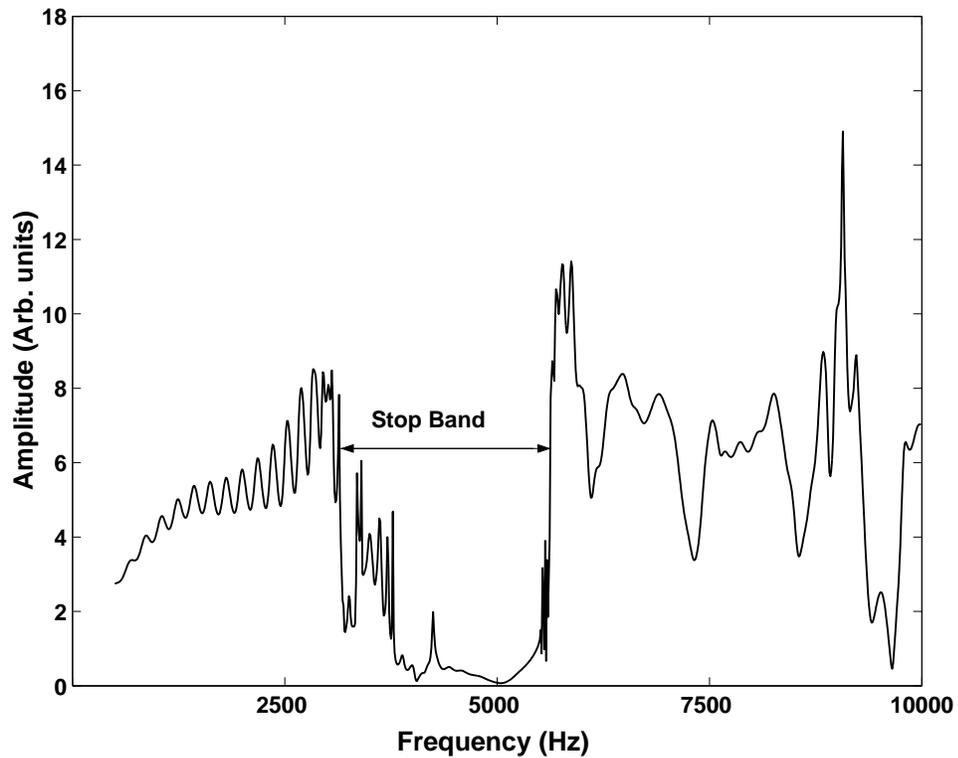} \vspace{12pt} \caption{
\label{fig1}\small Transmission as a function of $ka$ along the
$\Gamma$X (i.~e. [100]) direction for the square lattice.}
\end{center}
\end{figure}

The transmission spectrum for the square lattice of the rods is
presented in Fig.~\ref{fig1} for the propagation along the
$\Gamma$X (i.~e. [100]) direction. In the computation, we set the
number of cylinders to 200. The transmitter and receiver are
placed at such a small distance from the scattering array that the
boundary effects do not suppress the band gaps. Here we observe a
well defined inhibition regime ranging from about 3 kHz to 5.5
kHz. Within this range of frequency, the transmission is
significantly reduced. This agrees very well with the experimental
data shown in Fig.~3(a) of \cite{1998}. We have also performed a
series of numerical tests with respect to changing the number of
cylinders or the shape of the array. All results indicate that the
regime of inhibition is rather stable. For the transmission
outside this regime, however, the transmitted amplitude can vary
significantly as the number of scatterers or the shape of the
array changes. For example, the transmission through an array of
$10\times 20$ will differ from that through an array of $8\times
25$, even with the same lattice constant. The oscillatory
behaviour for frequencies below 2.8 kHz is purely caused by the
boundary. They may or may not appear, depending on the arrangement
of the array. But the inhibition behavior within the range between
3 and 5.5 kHz remains quantitatively the same for both arrays.
Such a stable inhibition regime is a clear indicator for the stop
band. This will be further confirmed by the band structure
calculation given below.

We also performed the transmission calculation using
eq.~(\ref{eq:final}) for propagation along the $\Gamma$M (i.~e.
[110]) direction. The band structure calculation indicates a small
stop band within about 5.6 and 6.0 kHz. Unfortunately, this stop
band cannot be clearly identified in our transmission calculation.
In the experiment, the transmission data in this case is also less
compelling\cite{1998}. The authors of \cite{1998} then used the
phase information extracted from the Fourier transformed data to
locate the anomalous phase delay caused by the stop band. Our
numerical data on the phase delay is again less convincing.
Several reasons may contribute to this. Among others, a prominent
reason may be due to the finite number of cylinders. In principle,
the stop band from the band structure calculation is obtained for
an infinitely large array of scatterers. The fact that there is
only a small gap in this situation would imply that a vastly large
number of scatterers is required in the transmission calculation.
Our present computing facilities, however, do not allow us to
simulate the scattering from an array of exceedingly large size.

\begin{figure}[hbt]
\begin{center}
\epsfxsize=5in \epsffile{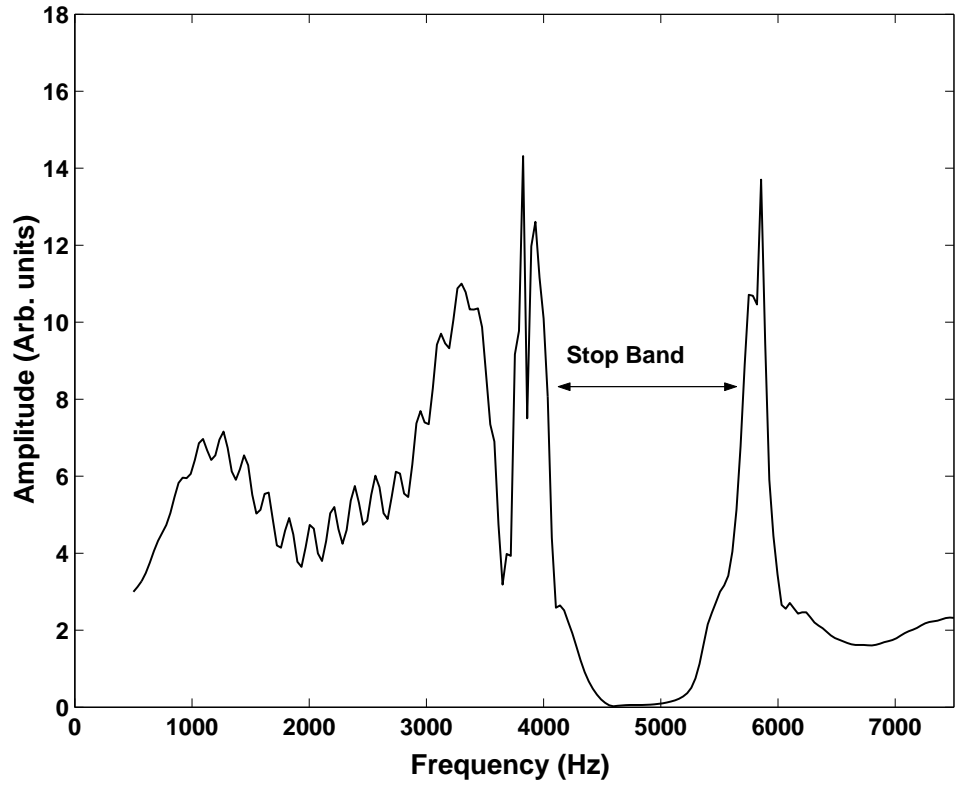}\vspace{12pt} \caption{
\label{fig2}\small Transmission as a function of $ka$ along the
$\Gamma$X (i.~e. [100]) direction for the triangular lattice.}
\end{center}
\end{figure}

We also performed numerical computation of the transmission
through the triangular lattice. The transmission spectrum is shown
in Fig.~\ref{fig2} for wave propagation along the $\Gamma$X
direction. Again we observe a stop band between 4 kHz and 5.7 kHz.
This is also in remarkable agreement with the experimental
observation, referring to Fig.~4(a) of \cite{1998}. The
computation of the transmission along the $\Gamma$M direction is
also done. Like in the case of the square lattice, the band
structure calculation indicates a narrow stop band within about
5.1 and 5.4 kHz. Again the stop band cannot be clearly identified
in our transmission calculation.

\begin{figure}[hbt]
\begin{center}
\epsfxsize=5in \epsffile{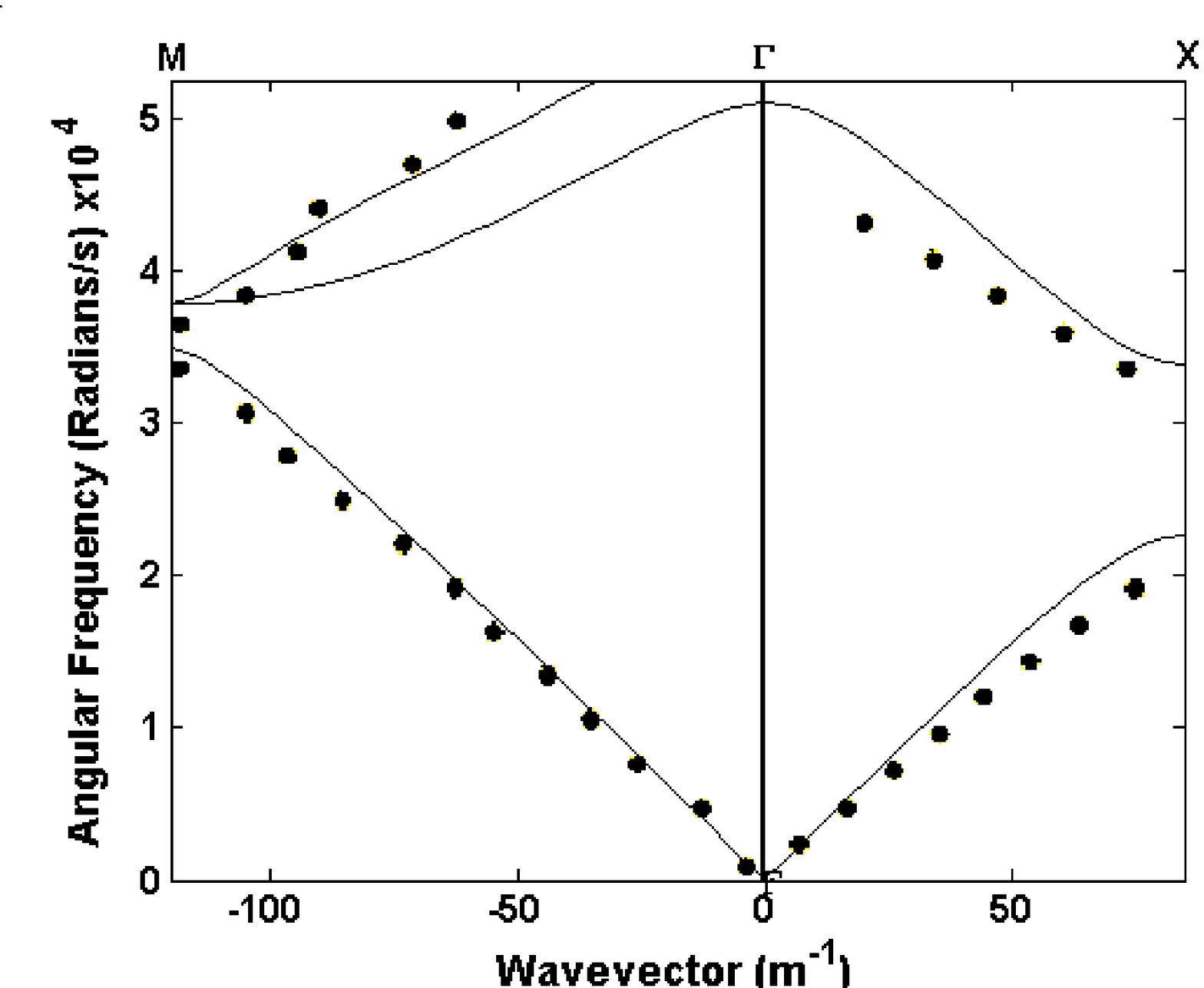} \vspace{12pt} \caption{
\label{fig3}\small The band structures computed by the plane wave
expansion method for the square lattice. The solid lines are the
theoretical results and the dots are the experimental data.}
\end{center}
\end{figure}

\begin{figure}[hbt]
\begin{center}
\epsfxsize=5in \epsffile{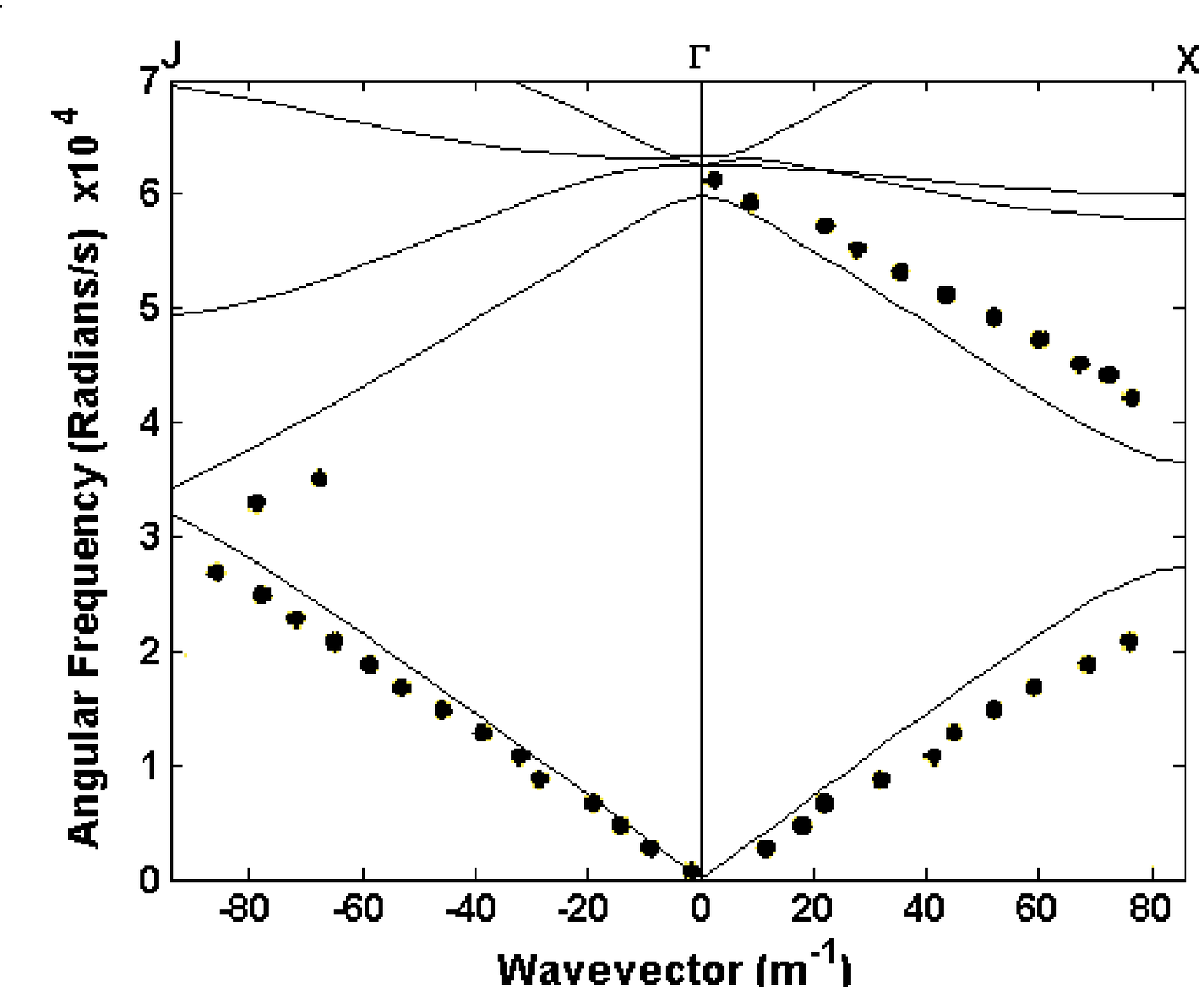} \vspace{12pt} \caption{
\label{fig4}\small The band structures computed by the plane wave
expansion method for the triangular lattice. The solid lines are
the theoretical results and the dots are the experimental data.}
\end{center}
\end{figure}

The theoretical dispersion relation is shown in Fig.~\ref{fig3}
and Fig.~\ref{fig4} for the square and triangular lattices
respectively. The experimental data read from \cite{1998} are also
plotted as the black dots in the figures.

First we consider the square lattice case. Overall speaking, the
experimental and theoretical data are in a good agreement. The
experimental data are only slightly lower than the theoretical
prediction. For propagation along the $\Gamma$M direction, the
experimental data agree remarkably well with the theory for the
first dispersion curve. For higher frequencies, the experimental
data seem to follow third dispersion band, though we see that some
experimental data fall on the second band. The agreement between
the theory and the experiment is slightly obscured by the presence
of three bands near the edge of the Brillouin zone. The theory
predicts a small band gap, as discussed earlier. Near the band
gap, the theory and the experiment are in a slight discrepancy.
Again, for the transmission along the $\Gamma$X direction, the
agreement in the lower dispersion band appears better than in the
higher band. A reason may be that the phase is relatively hard to
accurately measure at high frequencies. For both bands, the
experimental data are lower than the predicted values. Comparing
the band structure results in Fig.~\ref{fig3} with the
transmission results in Fig.~\ref{fig1}, we see that the stop band
predicted from the transmission spectrum is also slightly shifted
toward lower frequencies. The further computation indicates that
such a very small shift is due to the finite number of cylinders.
Increasing the number of scatterers will lift a bit the stop band
in the transmission spectrum.

Now we consider the triangular lattice. Again, from
Fig.~\ref{fig4} we see that the agreement between the theory and
the experiment is genuinely good, considering the complication
involved in the experiment. However, there are a few small
discrepancies. First, the experiment observes a wider stop band
along the $\Gamma$X direction, and the experimental observation of
the stop band along the $\Gamma$M direction is not so obvious.
From Fig.~\ref{fig4}, we see that the two lowest dispersion bands
are observed by experiment. In the triangular lattice case, the
stop band estimated from the transmission spectrum from
Fig.~\ref{fig2} agrees with the dispersion band calculation in the
$\Gamma$X direction.

\section{Summary}

In conclusion, in this paper we have presented a theoretical
analysis of the acoustic propagation through two-dimensional
regular arrays of parallel cylinders in air. A self-consistent
method is used to compute the wave transmission, taking into
account all orders of multiple scattering. We stress that this
approach in fact allows us to consider any configuration of the
scattering arrays. For the regular arrays, the plane-wave method
is used to calculate the band structures. Two lattice arrangements
are considered. The theory are then applied to the experimental
situations, yielding favorable agreements.

\begin{center}
{\bf Acknowledgments}
\end{center}
The work received support from National Science Council of ROC.


\input{epsf}

\newpage

\input{epsf}

\newpage

\input{epsf}

\newpage

\input{epsf}

\end{document}